\documentclass{llncs}

\usepackage[bookmarks=false]{hyperref}
\usepackage{amsmath}
\usepackage{listings}
\usepackage[all]{hypcap}
\usepackage{relsize}
\usepackage{xspace}
\usepackage{pgfplots}
\pgfplotsset{compat=1.14}

\newcommand{\inlcode}[1]{\texttt{#1}}
\newcommand{\Rplus}{\hspace{-.1em}\raisebox{.35ex}{\smaller{\smaller\textbf{+}}}}
\newcommand{\CC}{\mbox{C\Rplus\Rplus}\xspace}
\newcommand{\CS}{\mbox{C\raisebox{.35ex}{$\sharp$}}\xspace}

\begin{document}

\title{Performance Analysis of Zippers\thanks{This research was supported by SVV project number 260 453}}

\author{V\'it \v{S}efl}

\institute{
Faculty of Mathematics and Physics\\
Charles University, Prague, Czech Republic\\
\email{sefl@ksvi.mff.cuni.cz}
}

\maketitle

\begin{abstract}
A~zipper is a~powerful technique of representing a~purely functional data structure
in a~way that allows fast access to a~specific element. It is often used in cases
where the imperative data structures would use a~mutable pointer. However, the efficiency
of zippers as a~replacement for mutable pointers is not sufficiently explored. We attempt
to address this issue by comparing the performance of zippers and mutable
pointers in two common scenarios and three different languages: \CC, \CS, and
Haskell.
\end{abstract}

\section{Introduction}

Some programming techniques make use of the ability to keep a~pointer to internal
parts of a~data structure. Such a~pointer is usually called a~\emph{finger}
\cite{finger}. As an example, a~finger can be used to track the
most recently used node in a~tree. Tree operations can then start from the finger
instead of starting from the root of the tree, which can lead to a~speedup if the
program frequently operates on elements that are stored near each other.

However, fingers lose most of their utility when applied to purely functional data
structures. Operations that make use of fingers frequently require the structure to
contain pointers to parent nodes or require mutability. Pointers to parent nodes
create loops which hugely complicate update operations. 

A~\emph{zipper} \cite{huet} is a~technique of representing purely functional data
structure in a~way that allows direct access to an element at a~selected position.
Different data structures have different zipper representations: we, therefore,
distinguish between list zippers, tree zippers, etc. Zippers differ from fingers
in a~crucial way. Unlike a~finger, a~zipper contains the data structure. A~finger
can be removed, and the structure it was pointing to remains intact while removing
a~zipper removes the structure it contains. As a~consequence, while two fingers
give direct access to two positions, two zippers do not. 

Despite these differences, there is a~variety of tasks that can be solved by both
approaches. Our goal was to compare the effectiveness of these two techniques. We
chose two tasks where the ability to directly access a~position inside a~data
structure and perform local updates is beneficial: traversing a~tree in an arbitrary
way and building a~tree from a~sorted sequence. Each task was implemented in Haskell,
\CC, and \CS, using the programming style common to that language. Note that we
compared the performance difference between these techniques, rather than
performance across programming languages. 

This work is organized as follows. In the next section, we discuss zipper
representations. The third section looks at single position zippers in detail.
The testing methodology, as well as the programming tasks themselves, are presented
in the fourth section. Finally, the fifth section details our findings.

The source code used for performance testing is available online.
\footnote{\url{https://github.com/vituscze/performance-zippers}}

\section{Related Work}

Huet's original zipper technique \cite{huet} relies on manually analyzing the data
type and then defining the corresponding zipper structure. \autoref{lst:listzip}
shows an example of such a~zipper.

\begin{lstlisting}[caption={List and its zipper},label={lst:listzip},language=haskell]
data List a = Nil | Cons a (List a)

data ListZipper a = ListZipper
  { before :: List a
  , focus  :: a
  , after  :: List a
  }
\end{lstlisting}

This approach becomes problematic when working with heterogeneous data structures
(a~structure containing elements of multiple types), or when working with many
different zipper representations.

For heterogeneous collections, Huet's zipper can be used to represent only the
positions of one type of elements, which is quite limiting. Adams
\cite{scrapzippers} shows how to build a~zipper for heterogeneous collections by
using generic programming techniques based on the ideas of L\"ammel and Peyton Jones 
\cite{scrapboilerplate}. Another benefit of this approach is that new data
structures do not need a~custom implementation of the zipper structure, which
reduces the boilerplate that is usually present when dealing with zippers.

Instead of using an explicit data structure, the zipper can be represented as
a~suspended traversal of the original structure. Kiselyov \cite{oleg} uses
delimited continuations to implement suspended computation to great effect.
Applications include creating a~zipper for any type that is a~member of Haskell's
\inlcode{Traversable} type class, zipping two data structures for side-by-side
comparison and various operations on zippers capable of representing multiple positions.

Another way of dealing with the boilerplate code is to automate the generation of
auxiliary data structures. For each regular algebraic data type, the type of
one-hole contexts can be obtained by differentiating the original type, not unlike
differentiation in calculus \cite{derive,formalderive}. A~zipper is obtained by
combining an element of the original structure and the one-hole context. As a~result, the
zipper does not need to be defined for each data structure separately
\cite{typeindexeddata}. We explore this technique in more detail in the following
section.

Ramsey and Dias \cite{controlflow} use zippers to represent control flow
graphs in a~low-level optimizing compiler. The compiler is written in OCaml, giving the
opportunity to use an imperative approach based on mutable pointers as well as 
a~purely functional approach based on zippers. As part of their analysis, the authors
also include performance comparison. Zippers are shown to perform slightly better
than mutable pointers.

\section{Zipper}

Huet's zipper is based on the idea of pointer reversal. Reversing all pointers
along the path from the root of the structure to a~selected position called a~\emph{focus} 
creates a~structure that is rooted at the focus. This reversal has multiple advantages.
Direct access to the focus allows its modification in constant time. Even in
a~purely functional setting where in-place modifications are not available, creating
a~copy of the focused node may be used instead. The rest of the structure stays
intact and can be shared.

Similarly, accessing the parent and children of the focus can be done in
constant time, which can be used to efficiently move the focus around the
structure. Moving the focus is accomplished by reversing the pointers.

Huet shows how to represent this kind of pointer reversal as a~purely functional
structure. The nodes on the path from the root to the focus are stored in a~list.
Each element of the list must contain the values and substructures that are not
descended into as well as the direction taken when moving towards the focus. The
list is reversed, ensuring the parent of the focus is in the head position
(instead of the root of the structure).

\begin{lstlisting}[caption={Binary tree and its zipper},label={lst:treezip},language=haskell]
data Tree a = Leaf | Node (Tree a) a (Tree a)

data PathChoice a
  = NodeL a (Tree a) -- Focus is in the left subtree
  | NodeR (Tree a) a -- Focus is in the right subtree
    
data Context = Context
  (Tree a)       -- Left subtree of the focus
  (Tree a)       -- Right subtree of the focus
  [PathChoice a] -- Path to the root
    
data Zipper a = Zipper a (Context a)    
\end{lstlisting}

\autoref{lst:treezip} defines a binary tree and its zipper. \autoref{lst:focusmove}
shows how to move the focus of this zipper to the parent node.

\begin{lstlisting}[caption={Focus movement},label={lst:focusmove},language=haskell]
up :: Zipper a -> Maybe (Zipper a)
up (Zipper _ (Context _ _ [])) = Nothing
up (Zipper x (Context l r (NodeL p pr:ps))) = Just $
  Zipper p (Context (Node l x r) pr ps)
up (Zipper x (Context l r (NodeR pl p:ps))) = Just $
  Zipper p (Context pl (Node l x r) ps)
\end{lstlisting}

However, since the zipper structure depends on the original data structure, these
types and operations need to be defined for each structure separately. One way to
solve this problem is to automate this process by using data type differentiation
\cite{derive,formalderive}. We give a~brief overview of this technique here.

An \emph{algebraic data type} is a~data type defined as a~combination of products (tuples)
and sums (variants), potentially in a~recursive way. Algebraic data types that do
not change the parameters in recursive occurrences are known as \emph{regular types}. For
these types, the derivative is defined as follows.
\begin{align*}
\partial_x(0)& = 0 & \text{(empty type)}\\
\partial_x(1)& = 0 & \text{(unit type)}\\
\partial_x(y)& = 0 & \text{(type variable)} \\
\partial_x(x)& = 1 & \text{(type variable)} \\
\partial_x(F + G)& = \partial_x(F) + \partial_x(G) & \text{(sum type)}\\
\partial_x(F \times G)& = \partial_x(F) \times G +
  F \times \partial_x(G) & \text{(product type)}\\
\partial_x(\mu y. F)& = [\mu y. F/y] \partial_x(F) \times
  \text{List}\ ([\mu y. F/y] \partial_y(F)) & \text{(least fixed point)}
\end{align*}

The expression $[y/x]t$ denotes a~capture-avoiding substitution.
The variables can be introduced as parameters of the entire type (such as $a$ in $\text{List}\ a$)
or by the least fixed point operation, which is used to define recursive types.
The resulting derivative is a~type of \emph{one-hole contexts}. A~one-hole context is
a~structure that uniquely describes one position within the original data structure.
Zipper then consists of a~one-hole context together with an element of the original
structure.

For example, a~binary tree is a~regular algebraic data type, and its zipper can be
obtained by computing the derivative.
\begin{align*}
\partial_a(\text{Tree}\ a)
&= \partial_a(\mu x. 1 + x \times a \times x)\\
&= [\text{Tree}\ a/x]\partial_a(1 + x \times a \times x) \times
    \text{List}\ ([\text{Tree}\ a/x]\partial_x(1 + x \times a \times x))\\
&= [\text{Tree}\ a/x](x \times x) \times
    \text{List}\ ([\text{Tree}\ a/x](a \times x + x \times a))\\
&= \text{Tree}\ a \times \text{Tree}\ a \times
    \text{List}\ (a \times \text{Tree}\ a + \text{Tree}\ a \times a)
\end{align*}

This derivative matches the definition of the tree context given in
\autoref{lst:treezip}.

The zippers used for performance testing in this work were based on algebraic
data type differentiation. The resulting zipper representation was manually adjusted
to provide better control over its strictness properties.

\section{Performance Testing}

To compare the performance of zippers and fingers, we implemented tree traversal
and tree insertion in three different programming languages. The approach based
on zippers was implemented in Haskell. The approach based on fingers was implemented
in \CC and \CS. We included two imperative languages, one with manual memory
management and the other with garbage collection, to check how the memory management
model affected the relative performance. Unless specified otherwise, when discussing
the imperative solutions, we are talking about the \CC solution.

The tasks were chosen to test the performance under two different memory allocation
requirements. Tree traversal can avoid memory allocation altogether, while tree
insertion cannot. Both tasks were tailored to the finger- and zipper-based
approaches, which was done to better represent the common use case of these approaches.
In the following, we use the term \emph{cursor} to refer to either a~zipper or
a~finger.

\subsection{Tree Traversal}

The first task focuses on tree traversal. We are given a~binary tree and a~vector
describing positions within the tree together with replacement values. The goal is
to replace the specified elements of the original tree with the given values.

For cursor-based approach, the input vector contains instructions that specify
the movement of the cursor relative to its previous position. These movement
instructions are interspersed with the replacement instructions. The element
under the cursor is replaced with the given value whenever such instruction is
encountered. As an example, replacing the left child of the root with 10 and
the right child with 20 would be represented as
\inlcode{Vector.fromList [Mov~L, Set~10, Mov~U, Mov~R, Set~20]}.

We compared this approach to a~solution where the replacement operation always
starts at the root of the tree. The input vector describes the positions relative
to the root of the tree. When a~replacement value is encountered, the specified
element is replaced, and the position is reset back to the root of the tree. The
vector corresponding to the previous example would be \inlcode{Vector.fromList [Mov~L, Set~10, Mov~R, Set~20]}.
We do not allow \inlcode{Mov~U} as it is not necessary to describe a~position.

This input format was chosen for better control over the spatial locality of
the positions, which allowed us to observe how the cursor-based approach behaves
depending on the average distance between positions. This task also allowed us to
compare the performance of imperative solutions when memory allocation is not
a~factor.

\autoref{lst:ttspec} specifies the desired behavior of the root- and cursor-based
approaches. For simplicity, the specification does not handle incorrect inputs
(such as positions outside the tree).

\begin{lstlisting}[caption={Tree traversal specification},label={lst:ttspec},language=haskell]
data Tree a = Leaf | Node (Tree a) a (Tree a)
data Dir = L | R | U

-- Replace an element at position determined by a list
-- of left/right directions.
replace :: a -> [Dir] -> Tree a -> Tree a
replace v []     (Node l _ r) = Node l v r
replace v (L:ds) (Node l x r) = Node (replace v ds l) x r
replace v (R:ds) (Node l x r) = Node l x (replace v ds r)
replace _ _      t            = t

data Cmd a = Mov Dir | Set a

-- Specifies the behavior of the cursor-based approach.
cursor :: Tree a -> Vector (Cmd a) -> Tree a
cursor tree = fst . Vector.foldl step (tree, [])
  where
    step (t, ds) (Mov U) = (t, tail ds)
    step (t, ds) (Mov d) = (t, d:ds)
    step (t, ds) (Set v) = (replace v (reverse ds) t, ds)

-- Specifies the behavior of the root-based approach.    
root :: Tree a -> Vector (Cmd a) -> Tree a
root tree = fst . Vector.foldl step (tree, [])
  where
    step (t, ds) (Mov d) = (t, d:ds)
    step (t, ds) (Set v) = (replace v (reverse ds) t, [])    
\end{lstlisting}

\subsubsection{Imperative Solution}

\autoref{lst:ttlayout} defines the structures used to represent the binary
tree. Member functions are omitted for brevity.

\begin{lstlisting}[caption={Imperative binary tree (memory layout)},label={lst:ttlayout},language=c++]
struct node_t {
  node_t* parent;
  node_t* left;
  node_t* right;
  int64_t value;
};

struct tree_t {
  node_t* root;
  node_t* finger;  
};
\end{lstlisting}

Movement instructions are represented by integer constants to simplify the code.
The input vector is processed by iterating over all its elements, applying
the corresponding finger operation at each step. We evaluated the imperative
solutions on a~perfect binary tree of a~specified depth.

\subsubsection{Functional Solution}

The functional solution is more involved. Since the task is meant for a~cursor-based
approach, the zipper lends itself to this problem naturally. However, the root-based
approach presents a~few problems that have to be addressed.

The tree and zipper definitions shown in \autoref{lst:ttzip} follow the
definitions from \autoref{lst:treezip}, with the exception that each data type
contains strictness annotations. Fields annotated with \inlcode{!} are evaluated
whenever the enclosing data constructor is, which ensures that these structures
are fully evaluated at all times.

\begin{lstlisting}[caption={Binary tree and its zipper (with strictness annotations)},label={lst:ttzip},language=haskell]
data Tree = Node !Tree !Int64 !Tree | Leaf

data Path
  = PathLeft  !Int64 !Tree  !Path
  | PathRight !Tree  !Int64 !Path
  | Nil

data Zipper = Zipper !Tree !Int64 !Tree !Path
\end{lstlisting}

As a~consequence, the standard list type is replaced with a~custom type. GHC is also
instructed to unbox the integer fields, which is done to ensure that the cost of
operating on boxed values does not have any impact on the performance. Unboxed
vectors from the vector package are used to represent the input vector.

The zipper comes with operations that replace the focused element and move
the focus left, right, and up. Processing the input vector is implemented as
a~strict left fold. The zipper is the accumulator value, and in each step, we
apply zipper operation that corresponds to the element of the vector.

When starting from the root, replacing an element of the tree can be done easily
with a~recursive function that reads the vector in each recursive call and
descends into the correct subtree. The problem is propagating the information
about how many elements of the input vector were consumed so that the next
operation can start from the correct position. To make sure the root-based
approach is efficient, we compared a~few ways of dealing with this issue.

\paragraph{State Monad Solution}

The obvious solution is to use a~state monad. Note that laziness in the state is
unwanted, and the strict monad version is about twice as fast. Analyzing GHC's
core language \cite{ghccore}, the monadic code was optimized away, and most
values were unboxed. The only value that was not unboxed was the state returned
by the replacement operation. Replacing the standard state monad with
a~handwritten one that uses unboxed integer did not improve the performance in
a~statistically significant way, however.

\paragraph{ST Monad Solution}

Another way of passing the state is to use the imperative \inlcode{ST} monad.
The standard implementation of \inlcode{STRef} is limited to boxed types, which
hugely degraded the performance. The standard references had to be replaced with
unboxed references from the unboxed-ref package.

\paragraph{findIndices Solution}

Instead of propagating the new position via various versions of the state monad,
the replacement operation can be given hints on where to start. These hints can
be provided by an auxiliary vector containing the positions where each descent
starts. We can create this vector by using the \inlcode{findIndices} function
from the vector package. This solution has a~few issues. The input vector has to
be traversed twice, and the auxiliary vector has to be stored in the memory.

\paragraph{findIndex Solution}

We can avoid the memory allocation by computing the hints as needed, instead of
all at once, by using the \inlcode{findIndex} function.

\paragraph{Precomputed Vector Solution}

To measure the impact of the double traversal, we also implemented a~function
where the vector of hints is a~part of its input. The vector is precomputed, and
its time requirements were not included in the comparison.

Much like the imperative solution, all functional solutions were evaluated on
a~perfect binary tree of a~specified depth.

\subsection{Tree Insertion}

The second task focuses on tree building. Building a~search tree can be done much
more efficiently when the input sequence is sorted. The search for a~new insertion
point can be skipped since it will always be the leftmost or the rightmost node
(depending on the order of the input sequence).
This node can be tracked with a~finger that is updated
each time a~new element is inserted. The same can be done with a~zipper, although
the standard tree insert operation cannot be reused.

To test a~zipper for a~different structure, we chose 2-3 trees \cite{algo}
for this task. The structure is redundant: all data is kept in the leaf nodes,
and internal nodes contain the minimum of their right subtree (and of the middle
subtree, whenever applicable). The task is then to build a~redundant 2-3 tree
from a~descending sequence of a~given length. The standard approach starts
from the root of the tree when looking for the insertion point. The cursor-based
approach starts in the leftmost node and perform no additional search.

\subsubsection{Imperative Solution}

\autoref{lst:tilayout} defines the structures used to represent the 2-3 tree.
Member functions are omitted for brevity.

\begin{lstlisting}[caption={Imperative 2-3 tree (memory layout)},label={lst:tilayout},language=c++]
struct node_t {
  std::array<int64_t, 2> values;
  std::array<node_t*, 3> children;
  node_t* parent;
  bool is_two_node;
};

struct tree_t {
  node_t* root;
  node_t* last_inserted;
};
\end{lstlisting}

Tree insertion follows the standard algorithm. We obtain the insertion point
and attempt to insert the element into the corresponding leaf node. When the leaf
node is full, we allocate a~new node and redistribute all the elements from the
original node. After this split, we are left with a~two-node and a~three-node.
We take the middle element and the right node and attempt to insert them into the
parent node. We repeat this process until no split occurs or the root is reached.
Note that splitting an inner node results in two two-nodes because the middle
element does not need to be duplicated.

The split operation puts the inserted element into a~two-node when inserting
elements in descending order. As a~result, leaf nodes are only split every second
insertion. The implementation could be improved to also provide similar benefit
for insertion in ascending order.

We also tried the following variations of the tree operations: non-recursive
destructor, split operation that allocates the left node, and recursive
root-based insertion. The impact on the performance was either detrimental or
statistically insignificant.

We repeatedly inserted elements into the tree in descending order and measured the
time taken. In the case of \CC solution, this measurement also included the time
spent on deallocation, giving a~fairer comparison to the languages with
garbage collection.

\subsubsection{Functional Solution}

\autoref{lst:titree} shows a definition of 2-3 trees with strictness annotations.

\begin{lstlisting}[caption={Functional 2-3 tree},label={lst:titree},language=haskell]
data Tree
  = Leaf
  | Node2 !Tree !Int64 !Tree
  | Node3 !Tree !Int64 !Tree !Int64 !Tree
\end{lstlisting}

To insert an element into the tree, we recursively insert it into the correct subtree.
The result of this insertion is either one subtree or two subtrees and an element.
The first case is handled by replacing the corresponding subtree; the second case
indicates that a~split occurred and is handled similarly to the imperative solution.

To obtain a~zipper, we compute the derivative of a~parametrized version of the
2-3 tree type.

\begin{align*}
F &= 1 + a x^2 + a^2 x^3\\
\partial_a(F) &= x^2 + 2 a x^3\\
\partial_x(F) &= 2 a x + 3 a^2 x^2\\
\partial_a(\text{Tree}\ a)
&= \partial_a(\mu x. F)\\
&= [\text{Tree}\ a/x]\partial_a(F)
    \times \text{List}\ ([\text{Tree}\ a/x]\partial_x(F))\\
&= ((\text{Tree}\ a)^2 + 2 a (\text{Tree}\ a)^3)
    \times \text{List}\ (2 a (\text{Tree}\ a) + 3 a^2 (\text{Tree}\ a)^2)
\end{align*}

If the focus is in a~two-node, then there is only one choice for the position, and
the context is given by the two subtrees. This case is represented by
$(\text{Tree}\ a)^2$. If the focus is in a~three-node, there are two choices for
the position (left or right). The context is given by the three subtrees and the
element that is not focused, or $2 a (\text{Tree}\ a)^3$.

The path also distinguishes between two-nodes and three-nodes. In the case of a~two-node,
there are two choices for the focus position (left or right subtree). The context
is given by the element and the other subtree. This case is represented by
$2 a (\text{Tree}\ a)$. In the case of a~three-node, there are three choices for
the focus position (left, middle, or right subtree) and the context is given by
the two elements and the other two subtrees, resulting in the final term
$3 a^2 (\text{Tree}\ a)^2$.

Since the insertion algorithm only needs to know the leftmost node and not the
particular element, we simplify the zipper by removing this choice point. The
type variable is replaced with \inlcode{Int64} and the list type is replaced
with a~custom strict list. \autoref{lst:tizip} shows the resulting type.

\begin{lstlisting}[caption={2-3 tree zipper},label={lst:tizip},language=haskell]
data Nonempty
  = Nonempty2 !Tree !Int64 !Tree
  | Nonempty3 !Tree !Int64 !Tree !Int64 !Tree

data PathChoice
  = Path2L !Int64 !Tree
  | Path2R !Tree !Int64
  | Path3L !Int64 !Tree !Int64 !Tree
  | Path3M !Tree !Int64 !Int64 !Tree
  | Path3R !Tree !Int64 !Tree !Int64

data Path = Nil | Cons !PathChoice !Path
data Zipper = Zipper !Nonempty !Path  
\end{lstlisting}

Inserting an element by using a~zipper more closely resembles the imperative
solution. The key difference is that instead of pointers to parent nodes, the
zipper contains a~list of choices along the path from the root to the focus.
Instead of descending into the tree, the zipper-based insertion needs to descend
into this list.

When a~node splits and we attempt to add the element and one of the freshly split
nodes to the parent node, we also need to include information about the position
of the split node in relation to the element. This position is necessary to
reconstruct the extra information contained in the zipper. The imperative
solution assumes the split node is always to the right.

Much like the imperative solution, we repeatedly inserted elements into the tree in
descending order and measured time taken.

\section{Results}

All experiments were performed on Intel Core i7-4750HQ processor with 24 GB of main
memory under Windows 10 operating system. Each program was compiled with the
highest available level of compiler optimizations, and in the case of GHC,
LLVM backend was used for code generation. Garbage collectors were allowed to only
run in a~single thread. Each solution was executed with an increasing number of
iterations until a~time limit of three minutes was reached. The measured times were
normalized to one iteration. Mean execution time, as well as standard deviation,
were computed. Error bars represent one standard deviation. The raw measurements
are available online. \footnote{\url{https://github.com/vituscze/performance-zippers/blob/master/data.csv}}

\subsection{Tree Traversal}

The input files were generated by randomly picking 1,000,000 elements out of a~perfect binary
tree with 20 levels and outputting the path between them. We evaluated the tree
traversal in four scenarios which were obtained by biasing the random generator
towards particular areas of the tree: no bias, bottom bias, right bias, and bottom-right bias.
One input file was generated for each scenario to ensure any performance differences
were not due to different input data.

\begin{figure}
\centering
\begin{tikzpicture}
\begin{axis}[
  ybar,
  ymin=0, 
  enlarge x limits=0.2,  
  legend pos=outer north east,
  ymajorgrids,
  yminorgrids,
  minor grid style={line width=.1pt,draw=gray!20},
  minor y tick num=4,
  legend entries={State,ST,findIndices,findIndex,Precomputed},
  bar width=5pt,
  ylabel={Relative time (\%)},
  xtick=data,
  symbolic x coords={Uniform, Bottom, Right, Bottom-right},
  x tick label style={text width=40pt, align=center},
  width=\axisdefaultwidth,
  height=150pt
]
% State
\addplot [area legend,fill=red!30,error bars/.cd,y dir=both,y explicit] 
coordinates
{
(Uniform,129.01)
(Bottom,127.91)
(Right,184.83)
(Bottom-right,193.06)
};
% ST
\addplot [area legend,fill=orange!30,error bars/.cd,y dir=both,y explicit] 
coordinates
{
(Uniform,129.62)
(Bottom,127.58)
(Right,198.96)
(Bottom-right,208.25)
};
% Indices
\addplot [area legend,fill=yellow!30,error bars/.cd,y dir=both,y explicit] 
coordinates
{
(Uniform,100.97)
(Bottom,98.60)
(Right,122.80)
(Bottom-right,127.44)
};
% Index
\addplot [area legend,fill=green!30,error bars/.cd,y dir=both,y explicit] 
coordinates
{
(Uniform,100.00)
(Bottom,100.00)
(Right,100.00)
(Bottom-right,100.0)
};
% Precomputed
\addplot [area legend,fill=cyan!30,error bars/.cd,y dir=both,y explicit] 
coordinates
{
(Uniform,97.98)
(Bottom,97.40)
(Right,91.61)
(Bottom-right,94.75)
};

\end{axis}
\end{tikzpicture}
\caption{Tree Traversal Performance (Haskell)}
\label{fig:traversal-hs}
\end{figure}
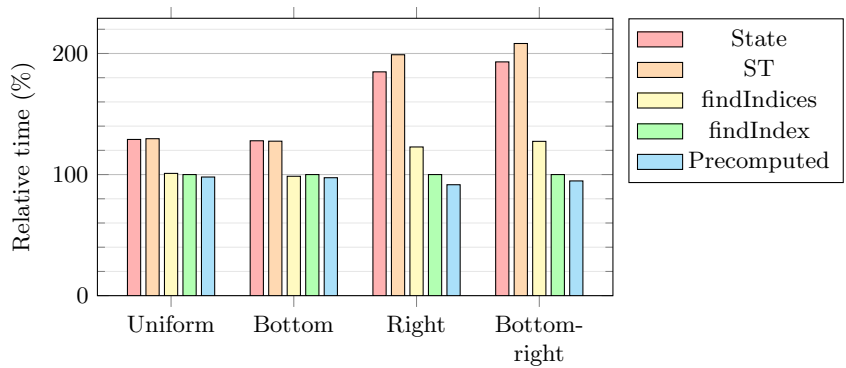

The results of the functional root-based approach are based on the
\inlcode{findIndex} solution. Its precomputed version is only marginally faster,
showing that the double traversal has a~low impact on the performance. The state
and \inlcode{ST} solutions are much slower. Interestingly, the \inlcode{ST}
solution is slightly slower than the purely functional state solution. Full
comparison of these variants can be found in \autoref{fig:traversal-hs}.

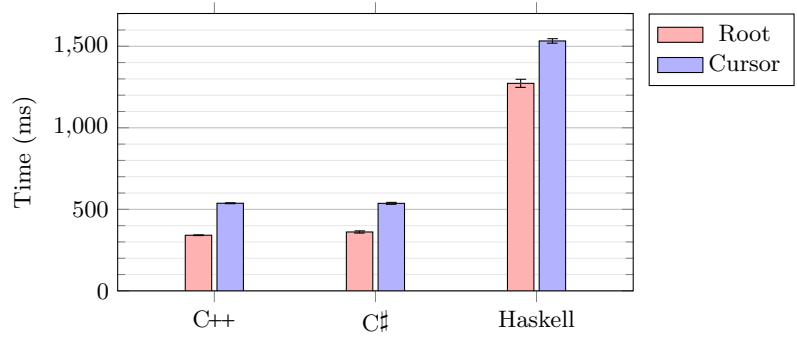
\begin{figure}
\centering
\begin{tikzpicture}
\begin{axis}[
  ybar,
  ymin=0,  
  enlarge x limits=0.3,  
  legend pos=outer north east,
  ymajorgrids,
  yminorgrids,
  minor grid style={line width=.1pt,draw=gray!20},
  minor y tick num=4,
  legend entries={Root,Cursor},
  bar width=10pt,
  legend columns=1,
  ylabel={Time (ms)},
  xtick=data,
  symbolic x coords={\CC,\CS,Haskell},
  width=\axisdefaultwidth,
  height=150pt
]
% Root
\addplot [area legend,fill=red!30,error bars/.cd,y dir=both,y explicit] 
coordinates
{
(\CC,341.16)+-(2.17,2.17)
(\CS,361.12)+-(6.78,6.78)
(Haskell,1272.65)+-(24.84,24.84)
};
% Cursor
\addplot [area legend,fill=blue!30,error bars/.cd,y dir=both,y explicit] 
coordinates
{
(\CC,537.31)+-(2.84,2.84)
(\CS,536.74)+-(5.93,5.93)
(Haskell,1532.65)+-(13.72,13.72)
};

\end{axis}
\end{tikzpicture}
\caption{Tree Traversal Performance (no bias)}
\label{fig:traversal-no}
\end{figure}

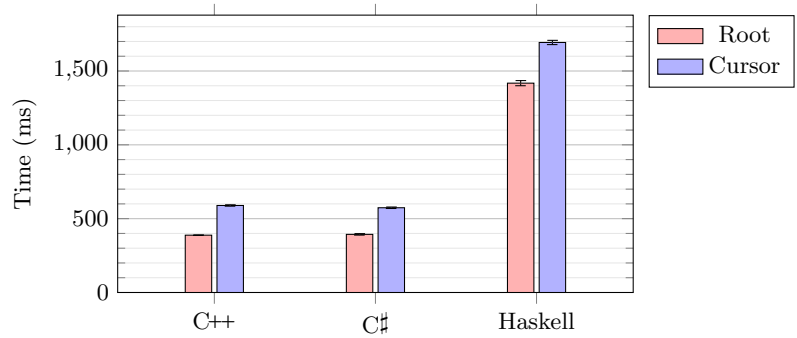
\begin{figure}
\centering
\begin{tikzpicture}
\begin{axis}[
  ybar,
  ymin=0,  
  enlarge x limits=0.3,  
  legend pos=outer north east,
  ymajorgrids,
  yminorgrids,
  minor grid style={line width=.1pt,draw=gray!20},
  minor y tick num=4,
  legend entries={Root,Cursor},
  bar width=10pt,
  legend columns=1,
  ylabel={Time (ms)},
  xtick=data,
  symbolic x coords={\CC,\CS,Haskell},
  width=\axisdefaultwidth,
  height=150pt
]
% Root
\addplot [area legend,fill=red!30,error bars/.cd,y dir=both,y explicit] 
coordinates
{
(\CC,388.17)+-(1.72,1.72)
(\CS,393.65)+-(5.02,5.02)
(Haskell,1417.70)+-(17.41,17.41)
};
% Cursor
\addplot [area legend,fill=blue!30,error bars/.cd,y dir=both,y explicit] 
coordinates
{
(\CC,589.31)+-(4.64,4.64)
(\CS,573.92)+-(4.98,4.98)
(Haskell,1693.21)+-(14.43,14.43)
};

\end{axis}
\end{tikzpicture}
\caption{Tree Traversal Performance (bottom bias)}
\label{fig:traversal-bottom}
\end{figure}

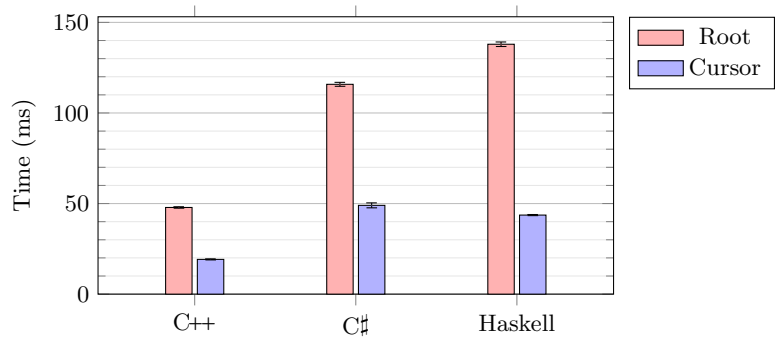
\begin{figure}
\centering
\begin{tikzpicture}
\begin{axis}[
  ybar,
  ymin=0,  
  enlarge x limits=0.3,  
  legend pos=outer north east,
  ymajorgrids,
  yminorgrids,
  minor grid style={line width=.1pt,draw=gray!20},
  minor y tick num=4,
  legend entries={Root,Cursor},
  bar width=10pt,
  legend columns=1,
  ylabel={Time (ms)},
  xtick=data,
  symbolic x coords={\CC,\CS,Haskell},
  width=\axisdefaultwidth,
  height=150pt
]
% Root
\addplot [area legend,fill=red!30,error bars/.cd,y dir=both,y explicit] 
coordinates
{
(\CC,47.84)+-(0.40,0.40)
(\CS,115.82)+-(1.09,1.09)
(Haskell,137.94)+-(1.25,1.25)
};
% Cursor
\addplot [area legend,fill=blue!30,error bars/.cd,y dir=both,y explicit] 
coordinates
{
(\CC,19.18)+-(0.29,0.29)
(\CS,49.01)+-(1.37,1.37)
(Haskell,43.64)+-(0.31,0.31)
};

\end{axis}
\end{tikzpicture}
\caption{Tree Traversal Performance (right bias)}
\label{fig:traversal-right}
\end{figure}

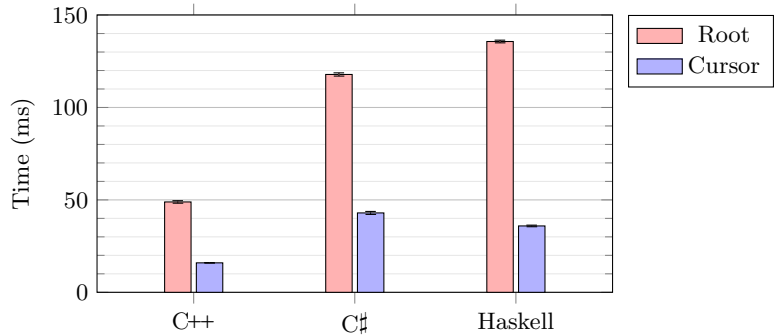
\begin{figure}
\centering
\begin{tikzpicture}
\begin{axis}[
  ybar,
  ymin=0,  
  enlarge x limits=0.3,  
  legend pos=outer north east,
  ymajorgrids,
  yminorgrids,
  minor grid style={line width=.1pt,draw=gray!20},
  minor y tick num=4,
  legend entries={Root,Cursor},
  bar width=10pt,
  legend columns=1,
  ylabel={Time (ms)},
  xtick=data,
  symbolic x coords={\CC,\CS,Haskell},
  width=\axisdefaultwidth,
  height=150pt
]
% Root
\addplot [area legend,fill=red!30,error bars/.cd,y dir=both,y explicit] 
coordinates
{
(\CC,48.89)+-(0.68,0.68)
(\CS,117.86)+-(0.88,0.88)
(Haskell,135.69)+-(0.73,0.73)
};
% Cursor
\addplot [area legend,fill=blue!30,error bars/.cd,y dir=both,y explicit] 
coordinates
{
(\CC,15.94)+-(0.14,0.14)
(\CS,42.95)+-(0.81,0.81)
(Haskell,35.91)+-(0.44,0.44)
};

\end{axis}
\end{tikzpicture}
\caption{Tree Traversal Performance (bottom-right bias)}
\label{fig:traversal-bottomright}
\end{figure}

When the spatial locality is low (\autoref{fig:traversal-no} and
\autoref{fig:traversal-bottom}), the root-based approach shows a~clear advantage
over the cursor-based approach. The relative gains of the root-based approach
are in the range of 50\% to 60\% for the imperative solutions and around 20\%
for the functional solution.

When the spatial locality is high (\autoref{fig:traversal-right} and
\autoref{fig:traversal-bottomright}), the cursor-based approach takes over. In
the case of the right bias, \CC reaches 150\% speedup, \CS 135\% and Haskell
220\%. Bottom-right bias increases this gap even more. \CC reaches 205\% speedup,
\CS 175\% and Haskell 280\%.

Notice that the root-based approach also shows a~considerable performance boost
when the input data has high spatial locality. This boost is a consequence of
cache-friendly memory access pattern. In all scenarios, the zipper-based
approach exhibits smaller performance losses (low spatial locality) and higher
performance gains (high spatial locality) when compared to the finger-based
approach.

\subsection{Tree Insertion}

Evaluating insertion into a~2-3 tree was done by repeatedly constructing a~tree
containing 10,000,000 elements. The ordered sequence was not part of the input.
Instead, the elements of this sequence were generated on the fly and inserted into the
tree directly, without any auxiliary structure. As mentioned earlier, this task
compared fingers and zippers in an environment where memory allocation is necessary.
For this reason, the \CC solution also evaluated the time it took to deallocate
the structure, giving a~better comparison with \CS and Haskell.

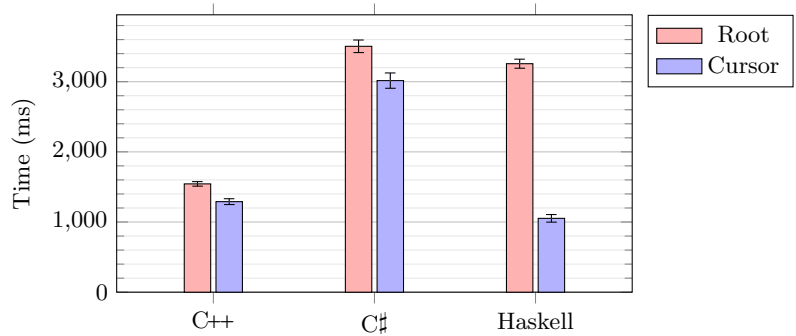
\begin{figure}
\centering
\begin{tikzpicture}
\begin{axis}[
  ybar,
  ymin=0,  
  enlarge x limits=0.3,  
  legend pos=outer north east,
  ymajorgrids,
  yminorgrids,
  minor grid style={line width=.1pt,draw=gray!20},
  minor y tick num=4,
  legend entries={Root,Cursor},
  bar width=10pt,
  legend columns=1,
  ylabel={Time (ms)},
  xtick=data,
  symbolic x coords={\CC,\CS,Haskell},
  width=\axisdefaultwidth,
  height=150pt
]
% Root
\addplot [area legend,fill=red!30,error bars/.cd,y dir=both,y explicit] 
coordinates
{
(\CC,1543.57)+-(32.32,32.32)
(\CS,3504.60)+-(89.59,89.59)
(Haskell,3256.66)+-(64.52,64.52)
};
% Cursor
\addplot [area legend,fill=blue!30,error bars/.cd,y dir=both,y explicit] 
coordinates
{
(\CC,1290.40)+-(41.57,41.57)
(\CS,3015.65)+-(109.26,109.26)
(Haskell,1052.39)+-(54.06,54.06)
};

\end{axis}
\end{tikzpicture}
\caption{2-3 Tree Insertion Performance}
\label{fig:insert}
\end{figure}

The results are shown in \autoref{fig:insert}. All three solutions show
a~preference for the cursor-based approach. In \CC and \CS, the finger-based
insertion is roughly 20\% faster than the root-based insertion. In Haskell,
the zipper-based insertion is 210\% faster.

Note that both the root-based and finger-based insertion allocate $\mathcal{O}(1)$
nodes (amortized) per insertion in imperative languages. The root-based
functional solution needs to copy the path from the root to the insertion point,
leading to $\mathcal{O}(\log{} n)$ new nodes per insertion. The zipper-based
insertion, therefore, not only avoids the cost of finding the insertion point
but also leads to significantly reduced allocation count.

Comparing the \CC and \CS results did not point to memory management as a major
factor. Reducing the size of the tree (by performing fewer insertions) showed that
the gap between \CC and \CS decreased slightly, which hints to a~minor performance
benefit from using garbage collection.

The \CC solution could be further optimized by using a memory pool instead of
the standard \inlcode{new} and \inlcode{delete} operators. However, we did not
want to deviate from the standard memory management models. In a similar vein,
we decided against fine-tuning the garbage collector parameters for the
Haskell and \CS solutions.

\section{Conclusion}

While zippers lack the flexibility and ease of use of mutable pointers, they are
nevertheless a~powerful tool when working with purely functional data
structures. However, it was unclear whether zippers offer the same performance
benefit as the imperative approach.

We compared fingers and zippers in two scenarios: arbitrary tree traversal
and tree insertion. The first test measured the effectiveness of zippers when its
imperative counterpart does not have to allocate memory. This test focused on
fast access to a~selected element as well as the ability to move the focus. The second
test considered the case where both the imperative and functional solutions need
to allocate memory. This test focused on the pointer reversal aspect of zippers.

We provided evidence that when zippers are used in a functional setting, they offer
higher performance gains compared to mutable pointers used in an imperative setting.
More importantly, zippers provide this gain without undermining the benefits of
purely functional data structures. We hope that this work encourages functional
programmers to use zippers before reaching for imperative techniques when optimizing
their code.

\clearpage

\bibliographystyle{splncs03}
\bibliography{zipper}

\end{document}